\documentstyle[12pt]{article}
\begin{document}
\begin{flushright}
{FAU-TP3-95/12}
\end{flushright}
\vspace*{2.0cm}
{\Large
\hspace*{2.5 cm}
{\bf QCD in the axial gauge}
}
\vskip 0.8cm
\hspace*{2.5 cm}
M. Thies
\footnote{Talk presented at the International School of Nuclear
Physics ``Quarks in Hadrons and Nuclei'', Erice, September 19-27, 1995}
\vskip 0.8cm
\hspace*{2.5 cm}
{\sl Institut f\"ur Theoretische Physik III, Universit\"at Erlangen\\
\hspace*{2.5 cm}
Staudtstr. 7, D-91058 Erlangen, Germany
}
\vskip 1.5cm

\vskip 0.7cm
\hspace*{2.5 cm}
{\bf ABSTRACT}
\vskip 0.5cm

We review a recent attempt to deal with non-perturbative features of
QCD by analytical means, using a manifestly gauge invariant, canonical
approach.

\setcounter{page}{1}
\setcounter{equation}{0}

\vskip 0.8cm
\hspace*{2.5 cm}
{\bf INTRODUCTION}
\vskip 0.5cm

Suppose you would like to solve the elementary quantum mechanical problem
of a particle in a central potential, but under the condition that only
$s$-waves are ``physical states'',
\begin{equation}
H=\frac{p^2}{2m}+V(r)\ , \qquad \vec{p} = \frac{1}{i} \frac{\partial}
{\partial \vec{r}} \ , \qquad [H, \vec{L}]= 0 \ ,
\label{1}
\end{equation}
\begin{equation}
\vec{L}|phys \rangle = 0 \ .
\label{2}
\end{equation}
This is a typical example of a constrained system, formulated in terms of
redundant variables. Here, the constraint can easily be resolved
by transforming to polar coordinates,
\begin{equation}
\left( -\frac{1}{2m} \frac{\partial^2}{\partial r^2} + V(r) \right)
u(r) = E u(r) \ , \qquad
u(r) = r\psi(r) \ .
\label{3}
\end{equation}
If we denote $r$ again by $x$, the result is suggestive of the
``axial gauge'' $y=z=0$. (Note however that the condition $x>0$ and
the boundary condition $u(0)=0$ for the radial wave function
are remnants of the transition to
curvilinear coordinates.)
Rotational symmetry is guaranteed, irrespective
of any further approximations
to dynamics, since we are using the scalar variable $r$.

In the case of gauge theories, we face a similar situation since
we start out with redundant variables. Can
we find the analogue of polar coordinates in QED or QCD, of course
with respect to local gauge symmetry
rather than rotational symmetry?
Let us consider QED first (Lenz {\em et al.}, 1994a). The
canonical formulation is most straightforward in the Weyl gauge ($A_0=0$)
since
$A_0$ has no conjugate momentum. All 3 spatial
components of $\vec{A}$ are then quantized. The Gauss law as
constraint on the
physical
states accounts for the fact that only 2/3 of the variables (the
transverse photons) are physical.
The Hamiltonian in the Weyl gauge reads
\begin{equation}
H=\int d^3x \left[ \psi^{\dagger}\left( -i \vec{\alpha} \vec{D}
+ \beta m \right) \psi + \frac{1}{2} \left( \vec{E}^2+ \vec{B}^2
\right) \right] \ ,
\label{4}
\end{equation}
with
\begin{equation}
\vec{D} = \vec{\nabla}-ie \vec{A} \ , \qquad \vec{E}(\vec{x})
= - \frac{1}{i} \frac{\delta}{\delta \vec{A}(\vec{x})} \ ,
\qquad \vec{B} = \vec{\nabla} \times \vec{A} \ .
\label{5}
\end{equation}
$H$ is invariant under local, time independent gauge transformations
\begin{equation}
\vec{A} \to \vec{A} + \vec{\nabla}\beta \ , \qquad
\psi \to e^{ie\beta}\psi     \ ,
\label{6}
\end{equation}
generated by the Gauss law operator
\begin{equation}
G(\vec{x}) = - \vec{\nabla}  \vec{E} + e \rho \ , \qquad
[G(\vec{x}),H] = 0 \ .
\label{7}
\end{equation}
Physical states are defined through the constraint
\begin{equation}
G(\vec{x})|phys \rangle = 0 \ .
\label{8}
\end{equation}
Here, unlike in the above toy model,
the choice of gauge invariant variables is far from unique, as could
have been guessed from
the proliferation of ``gauge choices'' in the
literature.
To see how one can resolve the constraint, write down the Gauss law in the
Schr\"odinger representation,
\begin{equation}
\left( \vec{\nabla} \frac{1}{i}\frac{\delta}{\delta \vec{A}} + e\rho
\right) \Psi[\vec{A},\psi] = 0 \ .
\label{9}
\end{equation}
This (functional) first-order differential equation can be solved
by a ``plane wave'' ansatz,
\begin{equation}
\Psi[\vec{A},\psi] = \exp \left\{ -i \int d^3x \vec{A}  \frac{1}{\Delta}
\vec{\nabla} e\rho \right\} \Phi[\vec{A}_{\rm tr},\psi ]
:= U^{\dagger} \Phi[\vec{A}_{\rm tr},\psi] \ .
\label{10}
\end{equation}
Since $U$ as defined in Eq. (\ref{10}) is a unitary operator, the
Schr\"odinger equation can be recast into the form
\begin{equation}
H\Psi = E \Psi \ \to \ U H U^{\dagger} \Phi = E \Phi \ .
\label{11}
\end{equation}
This is the equation corresponding to the radial Schr\"odinger
equation (\ref{3}) in the above example, provided we can show
that $UHU^{\dagger}$ contains only the physical variables
$\vec{A}_{\rm tr},\psi$. Indeed,
since the transformed Gauss operator
\begin{equation}
U G(\vec{x}) U^{\dagger} =  \vec{\nabla} \frac{1}{i}
\frac{\delta}{\delta \vec{A}(\vec{x})}
\label{12}
\end{equation}
commutes with $UHU^{\dagger}$, it is clear that the unitarily
transformed Hamiltonian cannot contain the longitudinal part of
$\vec{A}$ any more. A simple calculation confirms this expectation,
yielding
\begin{equation}
UHU^{\dagger} = \int d^3x \left[ \psi^{\dagger} \left( -i\vec{\alpha}
 \vec{D}_{\rm tr} + \beta m\right) \psi
+ \frac{1}{2} \left( \vec{E}_{\rm tr}^2+ \vec{B}^2 - e^2
\rho \frac{1}{\Delta} \rho \right) \right] \ .
\label{13}
\end{equation}
We have thus rederived the standard Coulomb gauge Hamiltonian,
including the familiar
static Coulomb potential. The Gauss law constraint is resolved, and all
redundant variables have been eliminated; in this quantum mechanical
scheme of gauge fixing, $U$ is denoted as
``unitary
gauge fixing transformation'' (Lenz {\em et al.}, 1994a).
In QED, the Coulomb gauge is
clearly singled out on physics grounds, since it is only in this gauge
that static sources decouple from the radiation field.
However, many other
gauges are conceivable; thus, for instance, the Gauss law can alternatively be
resolved
by the ansatz
\begin{equation}
\Psi[\vec{A},\psi] = \exp \left\{-i \int d^3x A_3
\frac{1}{\partial_3}\left(-\vec{\nabla}_{\bot}\vec{E}_{\bot}
+ e \rho \right) \right\} \Phi[\vec{A}_{\bot},\psi] \ .
\label{14}
\end{equation}
This leads to the axial gauge ($A_3=0$) which, however, is less
convenient from the
point of view of atomic or molecular
physics, since even static charges ``radiate''.

In QCD, it is much less obvious whether such a procedure can be carried
through and what the preferred gauge choice is (Lenz {\em et al.}, 1994b).
The Weyl gauge
Hamiltonian now reads
\begin{equation}
H=\int d^3x \left[ \psi^{\dagger} \left( -i\vec{\alpha} \vec{D}
+ \beta m \right) \psi + \mbox{tr}\left(\vec{E}^2+ \vec{B}^2\right)
\right] \ ,
\label{15}
\end{equation}
with the chromo-electric and -magnetic fields ($a=1,...,N_c^2-1$)
\begin{equation}
\vec{E}^a = - \frac{1}{i} \frac{\partial}{\partial \vec{A}^a} \ ,
\qquad \vec{B}^a = \vec{\nabla}\times \vec{A}^a + \frac{1}{2}
g f^{abc} \vec{A}^b \times \vec{A}^c \ .
\label{16}
\end{equation}
The Gauss law is non-linear,
\begin{equation}
(-\vec{D}\vec{E}+g\rho)|phys \rangle = 0 \ ,
\qquad (\vec{D}\vec{E})^a = \vec{\nabla}\vec{E}^a + gf^{abc}
\vec{A}^b\vec{E}^c \ ,
\label{17}
\end{equation}
hence it is much more difficult to resolve than the abelian one.
If we want to
implement the Coulomb gauge for instance, we have to solve
an equation of the type
($\vec{E}_{\ell}= \vec{\nabla} \phi$)
\begin{equation}
\vec{D}\vec{\nabla}\phi |phys  \rangle   = ... |phys \rangle
\label{18}
\end{equation}
for $\phi$. This is impossible (non-perturbatively) since one cannot
identify analytically the zero-modes
of the 2nd order partial differential
operator $\vec{D} \vec{\nabla}$ (Gribov, 1978).
By contrast, in order to implement the axial gauge, one
only
needs to solve a first order ordinary differential equation,
\begin{equation}
D_3E_3|phys \rangle = ... |phys \rangle   \ .
\label{19}
\end{equation}
Since one has control
over the zero modes of $D_3$,
it becomes possible to derive a Hamiltonian formulated exclusively
in terms of unconstrained,
physical variables in the axial gauge (Lenz {\em et al.}, 1994b).
The whole procedure is quite involved, and I will make no attempt to
go into any technical detail, nor even to show you the final Hamiltonian
in full glory.
Instead,
in the following
two sections, I will concentrate on one particularly instructive part
of the Hamiltonian, trying to exhibit some basic physics for which
the axial gauge is advantageous (Lenz {\em et al.}, 1995).

\vskip 0.8cm
\hspace*{2.5 cm}
{\bf CONFINEMENT}
\vskip 0.5cm

Let us come back to the axial gauge in QED for a moment and
point out one complication
which we have ignored so far. We work in a finite box with
periodic boundary conditions (i.e., on a ``torus''). In that case
$A_3=0$ is not a legitimate gauge,
simply
because the 2-dimensional field $a_3(\vec{x}_{\bot})=
\int_0^Ldz A_3(\vec{x}_{\bot},z)$
is gauge invariant. Physically, it describes transverse photons polarized
in
the 3-direction and propagating in the (1,2) plane.
In QED, this can easily be cured: Use the gauge $\partial_3A_3=0$
(rather than $A_3=0$)
by retaining $a_3(\vec{x}_{\bot})$ and eliminate some other variables instead
(Lenz {\em et al.}, 1994a).

In QCD, the corresponding
gauge invariant quantities formed exclusively out of $A_3$ are the
eigenvalues of
the spatial Wilson loop winding
around the torus (``Wilson line''),
\begin{equation}
\mbox{P} e^{ig\int_0^Ldz A_3} = V e^{iga_3L} V^{\dagger}
\label{20}
\end{equation}
($a_3$: diagonal matrix).
Here, the residual 2-dimensional variables cause much more trouble
than in QED.
In fact, most of
the work needed to resolve the non-abelian Gauss law has to do with
these ($N_{c}-1$) lower dimensional, color neutral fields, and
all the conspicuous
non-perturbative features displayed by the gauge fixed Hamiltonian
are somehow related to them (Lenz {\em et al.}, 1994b).
Among these features, most noteworthy is a Jacobian,
reflecting the transition from Lie algebra ($A_3$) to group elements (spatial
Wilson lines)
in the process of gauge fixing.

Is this a purely technical matter,
needed to properly define the theory in the infra-red, or
is there some real physics
associated with the $a_3$? By construction, it is clear that
the 2-dimensional variables $a_3$ are those whose
dynamics has been maximally simplified by the choice of the axial gauge;
as such,
they may teach us something about the dynamics of a whole class of
variables for which they are representative (but which are not
as simply described in our ``coordinate frame''). In order to exhibit
their physics content, let us do a very drastic approximation --
truncate the axial gauge Hamiltonian by keeping only
the terms in $a_3$ and the corresponding conjugate momenta
(Lenz {\em et al.}, 1995).
Surprisingly, even this primitive version of the Hamiltonian still
exhibits very interesting differences between QED and QCD, to which
we now turn.

Formally, the truncated Hamiltonian for both QED and SU(2) Yang Mills theory
reduces to the following 2-dimensional expression,
\begin{equation}
h=\int d^2x \left[ \frac{1}{2L} e_3^{\dagger}e_3 + \frac{L}{2}
(\vec{\nabla}_{\bot}a_3 )^2 \right] \ .
\label{21}
\end{equation}
Here, the electric field energy is found to be
\begin{equation}
e_3^{\dagger}e_3 = - \frac{1}{J} \frac{\delta}{\delta a_3} J
\frac{\delta}{\delta a_3}
\label{22}
\end{equation}
with the SU(2) Haar measure
\begin{equation}
J=\sin^2\left(\frac{gLa_3}{2}\right) \ ,
\label{23}
\end{equation}
whereas $J=1$ in the case of QED. This form
of the kinetic energy requires some ultra-violet regularization, for which we
choose
a transverse lattice (lattice spacing $\ell$, fundamental lattice
vectors $\vec{\delta}$).
(For QED, this would not be necessary,
but we do it anyway for ease of comparison.) Next, a ``radial'' wavefunctional
is introduced as $\tilde{\Psi}=\sqrt{J}\Psi$, and the kinetic
energy is reduced to standard form like in the above quantum mechanical
example, with a
corresponding
boundary condition ($\tilde{\Psi}=0$ whenever $J=0$). In terms of the
rescaled variable $\varphi=ga_3L/2$, the ``radial'' Hamiltonian then becomes
\begin{equation}
h=h_e+h_m = - \frac{g^2L}{8\ell^2}\sum_{\vec{r}} \frac{\partial^2}
{\partial \varphi_{\vec{r}}^2} + \frac{2}{g^2L} \sum_{\vec{r},\vec{\delta}}
\left(\varphi_{\vec{r}+\vec{\delta}}-\varphi_{\vec{r}}\right)^2 \ .
\label{24}
\end{equation}
Let us compare the abelian and non-abelian cases:
\vskip 0.1cm

i) QED: \\
Since the Jacobian is trivial, the Hamiltonian is quadratic and
can be
diagonalised by a standard lattice Fourier transform
($\varphi_{\vec{r}}=\frac{1}{N}\sum_{\vec{k}}e^{2\pi i
\vec{r}\vec{k}/N}\phi_{\vec{k}}$, with
$N=L/\ell$ the number of lattice sites in each
transverse direction.)
As a result of
the magnetic coupling, collective excitations appear, with
the dispersion relation
\begin{equation}
\omega_{\vec{k}}^2 = \frac{4}{\ell^2} \sum_{\vec{\delta}}
\sin^2 \left( \frac{\pi \vec{\delta}\vec{k}}{N} \right)
\to \left(\frac{2\pi\vec{k}}{L} \right)^2 \quad (|\vec{k}| \ll N) \ .
\label{25}
\end{equation}
In the limit $L \to \infty$, they become just
ordinary, massless photons.
The ground state wave functional is Gaussian,
\begin{equation}
\Psi \sim \prod_{\vec{k}} \exp \left(-\frac{4\ell^2}{g^2L_{\bot}}
\omega_{\vec{k}} \phi^{\dagger}_{\vec{k}} \phi_{\vec{k}} \right) \ ,
\label{26}
\end{equation}
and the virial theorem ensures that
fluctuations of electric and magnetic fields are equal,
\begin{equation}
\langle \vec{E}^2-\vec{B}^2 \rangle = 0  \ .
\label{27}
\end{equation}
Hence the axial gauge is not a bad choice at all for pure QED: We have
reduced
the problem of solving a 3-dimensional vector
field theory to that of a 2-dimensional
scalar field theory, by an appropriate choice of
coordinates. Although the Coulomb
gauge is preferred for static charges, the axial gauge is well suited
for studying the elementary gauge field excitations -- a transverse field is
naturally described in cartesian coordinates, the 3-axis pointing into
the direction of the polarization vector.
\vskip 0.1cm

ii) QCD:\\
Since $\tilde{\Psi}=0$ whenever $J=0$, we can restrict the variables
$\varphi_{\vec{r}}$ to the interval $[0,\pi]$. For large $L$, $h_e$
dominates over $h_m$, and we are left with the simple quantum mechanics
of infinite square well potentials at each site, totally decoupled
from each other.
The ground state is now uncorrelated in coordinate space,
\begin{equation}
\tilde{\Psi}_0 \sim \prod_{\vec{r}} \sqrt{\frac{L}{\pi}} \sin
\varphi_{\vec{r}} \ .
\label{28}
\end{equation}
(Note that this corresponds to the ``full'' wavefunctional $\Psi_0=
$const., $E_0=0$.) Since $\Psi_0$ is an eigenstate of the electric field
operator
with vanishing eigenvalue, we find trivially an exact ``dual Meissner effect''
\begin{equation}
\langle \vec{E}^2 \rangle = 0    \ .
\label{29}
\end{equation}
The vacuum has a magnetic condensate reminiscent of the
QCD vacuum,
\begin{equation}
\langle \vec{E}^2-\vec{B}^2 \rangle < 0 \ .
\label{30}
\end{equation}
Most importantly, there are no such excitations as ``plane wave gluons''; the
gap to the first excited state is
\begin{equation}
\Delta E = \frac{3g^2L}{8\ell^2} \to \infty \quad (L \to \infty) \ ,
\label{31}
\end{equation}
so that these gluonic degrees of freedom get frozen in the thermodynamic limit.
The magnetic contribution to the ground state energy can be evaluated
perturbatively; one finds
\begin{equation}
\langle 0 | h_m |0 \rangle = \frac{4L}{g^2 \ell^2}\left(\frac{\pi^2}{6}-1
\right) \ .
\label{32}
\end{equation}
In contrast to the QED case,
the vacuum has a precise value of the electric field (namely zero) and
therefore the
largest possible fluctuations in $\vec{A}$ (``stochastic''
vacuum). The excitation gap (\ref{31}) can be understood by comparing it to
lattice gauge theory: If one retains only $A_3$ and insists on Gauss's law,
the only gauge invariant excitations are
electric flux lines in the 3-direction around
the torus. Their energy in the strong coupling limit of Hamiltonian lattice
gauge
theory (Kogut and Susskind, 1975) is
\begin{equation}
\Delta E = \frac{g^2}{2\ell}\,j(j+1)\left( \frac{L}{\ell} \right)=
\frac{3g^2L}
{8\ell^2} \quad (j=1/2) .
\label{33}
\end{equation}
We thus recover the strong coupling string tension. Nevertheless,
our approach is quite different from the standard lattice where
one gets the same type of flux quantization and string tension also
in QED, in the strong coupling limit (Rothe, 1992).
Here, we obtain a qualitatively different behaviour in
the two cases, the difference being exclusively due to the Jacobian.

We have seen that
the simple degrees of freedom $a_3$ of the axial
gauge are useful in order to study the existence or non-existence
of plane wave gauge bosons.
We find strong indications that QCD does not
admit chromoelectric fields with constant polarisation vector over
large distances.
So far, we cannot say anything about the scale involved --
what is large? This would clearly require taking into account
the other gluonic degrees
of freedom as well and going through some renormalization procedure.
Nevertheless, we can get some information about the
relevant scale by indirect methods, using lattice results as input.
This is important in order to further assess the possible usefulness
of the axial gauge.

\vskip 0.8cm
\hspace*{2.5 cm}
{\bf DECONFINEMENT}
\vskip 0.5cm

It is generally believed that QCD undergoes a deconfining phase transition
to a quark gluon plasma at a temperature $T_c \simeq 150-200$ MeV
(M\"uller, 1985).
Can the axial
gauge Hamiltonian formulation of QCD contribute anything useful to
this issue? At first sight, the prospects look rather somber: Finite
temperature field
theory means compactification of the time direction, so that the Weyl
gauge $A_0=0$ is no longer allowed (for the same reason that the axial gauge
$A_3=0$ is not allowed on the torus). Moreover, there is evidence from
lattice gauge calculations that the spatial Wilson loops show area
law behaviour even above $T_c$ (see e.g. Karsch, 1994). This seems to
indicate that the ``simple'' variables
$a_3$ are not particularly sensitive to the phase transition, and that
consequently we would have no advantage over other approaches by
using the axial gauge.

Fortunately, nature provides us with a very elegant way out of these
problems. As a matter of fact, one can study thermodynamic properties like
the deconfining phase transition by working
strictly at $T=0$, but in a spatial
box contracted in one direction (say, the 3-direction) to
\begin{equation}
L_3=\beta=1/T \ .
\label{34}
\end{equation}
This result is intimately connected to Lorentz invariance (or, rather,
Euclidean O(4) invariance)
and would not
hold in non-relativistic field theories. Since it is rather unfamiliar,
let me
explain it with a simple example: Consider the partition function
\begin{equation}
Z = \int D[\phi] e^{-\int d^4x {\cal L}_E(x)}
\label{35}
\end{equation}
for a generic field theory in a finite box at finite temperature
in 2 cases:
\begin{eqnarray}
& I) & L_1,L_2,L_3 \gg \beta \ ,\nonumber \\
& II) & \beta,L_1,L_2 \gg L_3 \ .
\label{36}
\end{eqnarray}

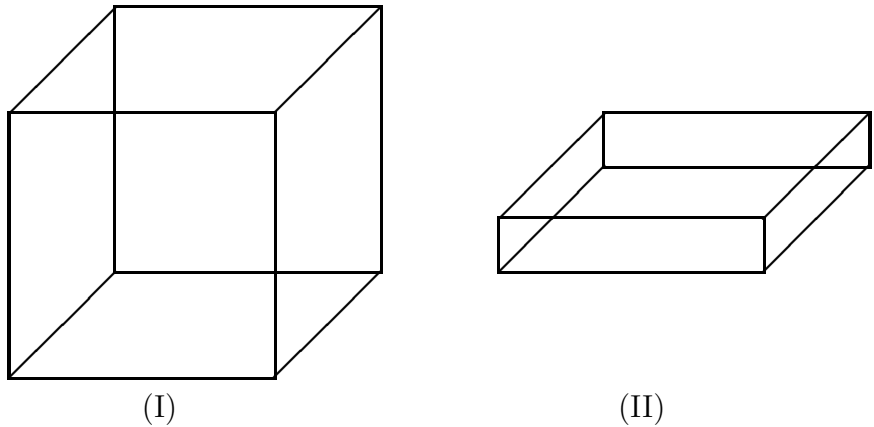
\thicklines
\begin{picture}(400,200)(00,0)
\put(75,30){\framebox(100,100)}
\put(115,70){\framebox(100,100)}
\put(75,30){\line(1,1){40}}
\put(175,30){\line(1,1){40}}
\put(75,130){\line(1,1){40}}
\put(175,130){\line(1,1){40}}
\put(125,15){(I)}

\put(260,70){\framebox(100,20)}
\put(300,110){\framebox(100,20)}
\put(260,70){\line(1,1){40}}
\put(360,70){\line(1,1){40}}
\put(260,90){\line(1,1){40}}
\put(360,90){\line(1,1){40}}
\put(305,15){(II)}

\end{picture}

Case (I) corresponds to a hot system in a large, isotropic box, case (II)
to a cold system in a box contracted along the 3-direction. Covariance
of the Euclidean action relates these 2 situations which
differ by the interchange of 2 coordinates, $x_3$ and the Euclidean time
$\tau$. Using the
standard relations
\begin{equation}
E=-\frac{\partial}{\partial \beta} \ln Z \ , \qquad
p = \frac{1}{\beta}\frac{\partial}{\partial V} \ln Z \ ,
\label{37}
\end{equation}
one finds for instance
\begin{equation}
\left( \frac{E}{V} \right)_I=-(p)_{II}\ ,
\label{38}
\end{equation}
and vice versa. These symmetry relations have surprising and powerful
consequences. By way of example,
consider the Casimir effect for a massless scalar field with periodic
boundary
conditions (Toms, 1980). If the system is enclosed between two plates at
distance
$L_3=d$,
there is an attractive force corresponding to a negative pressure
\begin{equation}
\label{39}
p=-\frac{\pi^2}{30} \frac{1}{d^4} \ .
\end{equation}
On the other hand, the energy density for an ultra-relativistic ideal
gas of scalar particles at finite
temperature is given by the Stefan Boltzmann law
\begin{equation}
\label{40}
\frac{E}{V} = \frac{\pi^2}{30} T^4 \ .
\end{equation}
These two laws of seemingly unrelated parts of physics are indeed
mapped onto each other
by the substitution $d \leftrightarrow 1/T$.
This observation -- which has been exploited several times in the literature
(cf. e.g. Toms, 1980; Koch {\em et al.}, 1992) --
opens a new perspective for studying thermodynamic
properties of field theories in a technically and conceptually
simpler way. All one has to do is to study the ground state in a
different geometry.

Once this is understood, it is not difficult to identify other relations
between at first sight
unrelated physical observables. One particularly amusing example
is the following: In (Lenz {\em et al.}, 1991), the Schwinger
model was studied
as a function of a parameter which measures how far off the light-cone
the quantization surface is. The fermion condensate was evaluated
analytically as a function of this parameter. It was also pointed out
that the same function
can be interpreted as dependence of the condensate on the size
of the box. Independently, the finite temperature Schwinger model was
investigated in (Sachs and Wipf, 1992), and again the fermion condensate
evaluated in closed form
as function of $T$. The two formulae (Eq. (3.105) of
(Lenz {\em et al.}, 1991)
and Eq. (5.10) of (Sachs and Wipf, 1992))
agree exactly if one identifies
the corresponding variables,
\begin{equation}
\eta_g' = \frac{\pi}{2} (\beta m_{\gamma})^2 \ ,
\label{41}
\end{equation}
a fact which seems to have been overlooked so far.
Similarly, it is tantalizing to reinterpret corresponding findings
for large $N_c$ QCD$_2$ in (Lenz {\em et al.}, 1991) as
evidence for a chiral symmetry
restoring phase transition at finite temperature, in contradiction to the
common lore (McLerran and Sen, 1985; Ming Li, 1986).

Let us now return to axial gauge QED and QCD. We
start with QED and go one step beyond the truncation of $H$, taking into
account perturbatively the coupling of $a_3$ to the charged fermions.
The simplest way to do this is to
compute the effective potential for the zero mode of $a_3$
by evaluating the energy density of the Dirac sea in a constant
background potential $a_3$.
In view of the application to finite temperature field theory,
we have to require anti-periodic boundary conditions
for the fermions,
\begin{equation}
\psi(\vec{x}_{\bot},L) = - \psi(\vec{x}_{\bot},0) \ .
\label{42}
\end{equation}
We can gauge away a constant $a_3$, provided we change these boundary
conditions into quasi-periodic ones,
\begin{equation}
\psi(\vec{x}_{\bot},L) = - e^{iea_3L} \psi(\vec{x}_{\bot},0) \ .
\label{43}
\end{equation}
This yields the following discretization for the 3-component of the
fermion momenta,
\begin{equation}
p_3 = \frac{\pi}{L}(2n+1)-ea_3 \ .
\label{44}
\end{equation}
The effective potential is then given by
\begin{equation}
U_{\rm eff}(a_3) = -2 U(c)
\label{45}
\end{equation}
with $U(c)$ the (heat-kernel regularized) sum over single particle
energies,
\begin{equation}
U(c) = \lim_{\lambda \to 0} \frac{1}{L}\int
\frac{d^2p_{\bot}}{(2\pi)^2}
\sum_{n=-\infty}^{\infty} E(\vec{p}_{\bot},n-c) e^{-\lambda
E(\vec{p}_{\bot},n-c)}    \ ,
\label{46}
\end{equation}
The variable $c$ is defined as
\begin{equation}
c= \frac{eLa_3}{2\pi} - \frac{1}{2} \ ,
\label{47}
\end{equation}
and
\begin{equation}
E(\vec{p}_{\bot},\nu) = \sqrt{p_{\bot}^2+ ( 2\pi\nu/L)^2}\ .
\label{48}
\end{equation}
Performing the integral in (\ref{46}) and using the generating
function for Bernoulli polynomials, one finds
\begin{eqnarray}
U(c) & = & \lim_{\lambda \to 0} \frac{1}{2\pi L} \frac{\partial^2}
{\partial \lambda^2} \frac{1}{\lambda} \sum_{n=-\infty}^{\infty}
e^{-2\pi\lambda |n-c|/L}
\nonumber \\
& = & \frac{2\pi^2}{3 L^4} B_4(c)
\label{49}
\end{eqnarray}
(valid for $|c|<1$, and to be continued periodically outside this
interval).
We have dropped a $c$-independent, infinite constant. The
result for the effective potential is therefore
\begin{eqnarray}
U_{\rm eff}(a_3) &=& -\frac{4\pi^2}{3L^4} B_4(c) \nonumber \\
& = & -\frac{4\pi^2}{3L^4}\left( c^2 (1-c)^2- \frac{1}{30}
\right)
\label{50}
\end{eqnarray}
$U_{\rm eff}$ has extrema at $c=0,\frac{1}{2},1$, the minimum
corresponding to
$c=\frac{1}{2}$
or $a_3=\frac{2\pi}{eL}$.
Expanding $U_{\rm eff}$ to 2nd order
around the minimum, we find
\begin{equation}
U_{\rm eff}\left(\frac{2\pi}{eL} + \delta a_3 \right) \simeq
- \frac{7}{4} \left( \frac{\pi^2}{45L^4} \right) + \frac{1}{2} \left(
\frac{e^2}{3L^2}\right) \delta a_3^2 \ .
\label{51}
\end{equation}
{}From this expression, replacing $L$ by $\beta= 1/T$, we can read off
the familiar free energy density of an ideal gas of massless fermions,
\begin{equation}
f = - \frac{7}{4} \left( \frac{\pi^2 T^4}{45} \right) \ ,
\label{52}
\end{equation}
as well as the ``electric mass'' of the photon (Kapusta, 1989),
\begin{equation}
m_{\rm el}^2 = \frac{1}{3} e^2T^2 \ ,
\label{53}
\end{equation}
in spite of the fact that we have only been dealing with ground state
properties.
This gives a first impression of the potential use of the axial gauge,
if at the same time one reinterprets
the contracted box in terms of finite temperature
field theory.

Let us now turn to the corresponding computations for QCD. Here,
if one interchanges 3- and time directions, the
simple variables $a_3$ in the axial gauge become the eigenvalues of the
thermal Wilson lines, the standard order parameters for the
confinement-deconfinement transition. This is obviously a welcome
feature which means that we have chosen a useful set of variables and are in
a good position to derive an effective theory for this order parameter
in the spirit of Landau Ginzburg theory.

Ignoring quarks (which could be taken into account easily as
well in the present approximation),
we evaluate
the zero-point energy of the
``perpendicular'' gluons $A_1, A_2$ in the presence of a constant
background field $a_3$. We treat the $\vec{A}_{\bot}$ as free fields, except
for the minimal coupling to $a_3$. Again, if $a_3$ is constant, its effect is
equivalent to changing the boundary conditions in 3-direction from
periodic to quasi-periodic ones.
The expression for the vacuum
energy density can be written down in complete analogy to the QED example.
We consider SU(2) Yang-Mills theory and introduce the rescaled
field variable
\begin{equation}
c=\frac{gL}{2\pi} a_3  \in [0,1]   \ .
\label{55}
\end{equation}
The vacuum energy density of $\vec{A}_{\bot}$ will serve as
effective potential for $a_3$. It can be decomposed as
\begin{equation}
U_{\rm eff} = 2 U(c) + U(0) \ ,
\label{56}
\end{equation}
where the two different terms come from the (color) charged and neutral
components
of $\vec{A}_{\bot}$,
respectively. The function $U(c)$ is exactly the same as the one
given in Eq. (\ref{46}).
This yields for $U_{\rm eff}$, Eq. (\ref{56}), the form
\begin{equation}
U_{\rm eff} = \frac{4\pi^2}{3L^4}\left( c^2 (1-c)^2- \frac{1}{20}
\right)
\label{57}
\end{equation}
(here again, this function has to be continued periodically
outside of the interval $0 \leq c \leq 1$). The value of $U_{\rm eff}$
in the minima
reproduces the free energy density of an ideal gluon gas,
if we replace $L$ by $\beta=1/T$,
\begin{equation}
U_{\rm eff}|_{c = 0} = -\frac{\pi^2}{15 L^4} \to - \frac{\pi^2}
{45} T^4 (N_c^2-1) \quad (N_c=2) \ .
\label{58}
\end{equation}
Using the same substitution,
we reproduce the ``one-loop effective
potential'' for the ($\vec{x}$-independent) thermal Wilson line, which has
been discussed extensively in the literature
(first for SU(2) by Weiss, 1981).
A look at the original reference shows that our derivation is significantly
simpler.

In the standard approach to finite temperature QCD, one infers
the gluon electric mass from the
2nd derivative of $U_{\rm eff}$ in a minimum,
\begin{equation}
m_{\rm el}^2 = \left. \frac{\partial^2 U}{\partial a_3^2} \right|_{c=0}
= \frac{2g^2}{3L^2} \to \frac{1}{3}N_c g^2 T^2
\label{59}
\end{equation}
{}From this point of view,
it would seem that QED and QCD behave quite similarly indeed.
However, so far, we have only discussed the effective potentials and
ignored the difference in the kinetic energy of $a_3$, which was crucial
to ``confine'' plane wave gluons. In our approach, we see no justification
for disregarding these effects, which are overwhelming at low temperature.

Let us try to
understand at least qualitatively the effect of the Jacobian.
Following (Polchinski, 1992), we first assume
that the effective potential derived
for a constant $a_3$ can be taken over for $\vec{x}_{\bot}$-dependent
$a_3$ as well, with the same functional form.
The effective potential to be added to $h_e+h_m$, Eq. (\ref{24}), is then
(up to
an irrelevant constant term)
\begin{equation}
u_{\rm eff} = \frac{4\ell^2}{3\pi^2 L^3} \sum_{\vec{r}} \varphi_{\vec{r}}^2
\left( \pi - \varphi_{\vec{r}} \right)^2 \ .
\label{60}
\end{equation}
Secondly, we model the effect of the Jacobian by approximating the
infinite square well potential by a harmonic oscillator potential, chosen such
as to reproduce the gap between ground state and first excited state,
eq. (\ref{31}),
\begin{equation}
u_{\rm conf} = \frac{9 g^2L}{32 \ell^2} \sum_{\vec{r}} \left(
\varphi_{\vec{r}}- \pi/2 \right)^2 \ .
\label{61}
\end{equation}
The resulting potential $u_{\rm tot}= u_{\rm eff}+u_{\rm conf}$ now
exhibits the correct $Z_2$ ``center symmetry'', whereas without the
confining potential it has a discrete translational symmetry both
in QED and QCD. With a quartic potential and a quadratic one (possibly
changing sign),
\begin{equation}
u_{\rm tot}= \left( \frac{9g^2L}{32 \ell^2}-\frac{2\ell^2}
{3L^3} \right) \sum_{\vec{r}} \left( \varphi_{\vec{r}}-\pi/2 \right)^2
+ \frac{4\ell^2}{3\pi^2 L^3} \sum_{\vec{r}} \left( \varphi_{\vec{r}}
-\pi/2 \right)^4 + \mbox{const.} \ ,
\label{62}
\end{equation}
we are now in the standard situation for a second order
phase transition and can determine the gluon mass below and above
the critical temperature. It is convenient to introduce 2 masses,
the electric gluon mass of Eq. (\ref{59}) and the ``confining mass''
of Eq. (\ref{31}),
\begin{equation}
m_{\rm el}^2 = \frac{2g^2}{3L^2} \ , \quad m_{\rm conf}^2
= \left( \frac{3g^2L}{8\ell^2} \right)^2 \ .
\label{63}
\end{equation}
The critical temperature (using $L=1/T$)
where the quadratic term in $u_{\rm tot}$ vanishes is then
determined by the condition
$m_{\rm conf}^2- m_{\rm el}^2/2 = 0$, or
\begin{equation}
T_c^4 =  \frac{27 g^2}{64 \ell^4}               \ .
\label{64}
\end{equation}
The expressions for the gluon mass in the confined and deconfined
phase are $m_{\rm eff}^2 = m_{\rm conf}^2 - m_{\rm el}^2/2 $  and
$m_{\rm eff}^2 = m_{\rm el}^2- 2 m_{\rm conf}^2 $,
respectively. These results are by no means
realistic, but are reminiscent of discussions of
the phase transition in the strong coupling limit (Gross, 1983).
Nevertheless,
it is interesting that the crudest conceivable approximation to the
gauge fixed Hamiltonian has already the potential of describing
both confinement and deconfinement of gluons. Much more work is needed
in order to properly understand how the confining effects (related to
the Jacobian) are overcome at high temperature and how the well
established perturbative features of QCD can be recovered in
the present framework.

We finish with
the following simple observation, which is again related to the question
of length scales.
If we translate the
critical temperature as determined from lattice gauge calculations
(Karsch, 1994)
into
a critical length, we learn that the phase transition to the quark gluon
plasma takes place if we contract space in 3-direction to
\begin{equation}
L_c=1/T_c \simeq 1.0-1.3 \ \mbox{fm} \ .
\label{54}
\end{equation}
Thus, the variables $a_3$ should be qualitatively representative for
a large class of waves indeed -- all those whose polarization vector
does not change over distances appreciably larger than 1 fm.
On the other hand, $L_c$ also sets the scale for
the length at which quark and gluon degrees of freedom become
essential -- a rather large value from the point of view of nuclear physics.

\vskip 1.8cm
\hspace*{2.5 cm}
{\bf ACKNOWLEDGEMENT}
\vskip 0.5cm
This talk is based on work done in collaboration with F. Lenz, E.J. Moniz,
H.W.L. Naus,
K. Ohta, and M. Shifman.
I am grateful to H. Grie\ss hammer for critically reading the manuscript.
I would like to thank the organizers of the 1995 Erice School on Nuclear
Physics, in particular
Prof. A. Faessler, and the Deutsche Forschungsgemeinschaft (DFG),
for the opportunity to present this contribution and for financial support.

\vskip 1.8cm
\hspace*{2.5 cm}
{\bf REFERENCES}
\vskip 0.5cm

V.N. Gribov, (1978). {\sl Nucl. Phys.}, {\bf B139}, 1.

M. Gross, (1983). {\sl Phys. Lett.}, {\bf B132}, 125.

J.I. Kapusta, (1989). {\sl "Finite Temperature Field Theory"}, Cambridge,
chapter 5.

F. Karsch, (1994). {\sl Nucl. Phys. (Proc. Suppl.)}, {\bf B34}, 63.

V. Koch, E.V. Shuryak, G.E. Brown, and A.D. Jackson, (1992).
{\sl Phys. Rev.}, {\bf D46}, 3169.

J. Kogut and L. Susskind, (1975). {\sl Phys. Rev.}, {\bf D11}, 395.

F. Lenz, M. Thies, S. Levit, and K. Yazaki, (1991). {\sl Ann. Phys.},
{\bf 208}, 1.

F. Lenz, H.W.L. Naus, K. Ohta, and M. Thies, (1994a). {\sl Ann. Phys.},
{\bf 233}, 17.

F. Lenz, H.W.L. Naus, and M. Thies, (1994b). {\sl Ann. Phys.},
{\bf 233}, 317.

F. Lenz, E.J. Moniz, and M. Thies, (1995).  {\sl Ann. Phys.}, {\bf 242}, 429.

Ming Li, (1986). {\sl Phys. Rev.}, {\bf D34}, 3888.

L.D. McLerran and A. Sen, (1985). {\sl Phys. Rev.}, {\bf D32}, 2794.

B. M\"uller, (1985). {\sl "The Physics of the Quark-Gluon Plasma"},
Springer, p. 27.

J. Polchinski, (1992).
{\sl Phys. Rev. Lett.}, {\bf 68}, 1267.

H.J. Rothe, (1992).
{\sl "Lattice Gauge Theories"}, World Scientific, Singapore, ch. 10.

I. Sachs and A. Wipf, (1992). {\sl Helv. Phys. Acta}, {\bf 65}, 652.

D.J. Toms, (1980). {\sl Phys. Rev.}, {\bf D21}, 928.

N. Weiss, (1981). {\sl Phys. Rev.}, {\bf D24}, 475.

\end{document}